\documentclass[journal]{IEEEtran}

\usepackage[pdftex]{graphicx}
\graphicspath{{Figures/}}
\DeclareGraphicsExtensions{.pdf,.jpeg,.png}

\usepackage{amsmath}
\usepackage{graphicx}
\usepackage{enumerate}
\usepackage{url}
\usepackage{subfigure}
\usepackage{algorithm}
\usepackage{algpseudocode}
\usepackage{stfloats}
\usepackage{lineno}
\usepackage{hyperref}
\usepackage{bm}
\usepackage{bbm}
\usepackage{enumerate}
\usepackage{xcolor}
\usepackage{float}

\newtheorem{remark}{Remark}

\newtheorem{philosophy}{Philosophy}
\newtheorem{axiom}{Axiom}

\begin{document}
%
\title{Thinking towards Motion Modeling: on High-order Derivatives of Displacement}

\author{Shixiong~Wang 
        and Andrew~Lim
\thanks{S. Wang is the Corresponding Author.}
\thanks{S. Wang, and A. Lim are with Department of Industrial Systems Engineering and Management, National University of Singapore. E-mail: s.wang@u.nus.edu; isealim@nus.edu.sg.}
}

\maketitle

\begin{abstract}
This note is concerned with the problem of motion modeling in Physics. We aim to use the Time-Variant Local Autocorrelation Polynomial to understand the mathematical model of motion description. The mathematical explanation of existences of physical concepts, beyond Velocity and Acceleration, like Jerk, Snap, Crackle, and Pop could be revealed.
\end{abstract}


\IEEEpeerreviewmaketitle

\section{Problem Statement}  \label{sec:introduction}
In literature like \cite{eager2016beyond,thompson2011snap}, the concepts of Jerk, Snap, Crackle and Pop, beyond Velocity and Acceleration, are reported. Inspired by this, we are interested in describing the motion with mathematical model. That is, we aim to find the connections among the displacement and other physical quantities.

\section{Polynomial Regression of a Continuous Function}
This section, before we investigate the motion modeling, introduces some basis theories of function approximation.

Suppose we have a continuous function $f(t)$, where $t$ denotes the continuous time variable. Let $n$ denote the discrete time index. Thus we have $t := nT_s$ where $T_s$ is the time gap between $n$ and $n+1$ (for notation simplicity, we interchangeably use $T$ and $T_s$). We aim to find the best fitting of $f(t)$.

One available way is with polynomial regression.

The theoretical validity and sufficiency of polynomial regression is from the prestigious Weierstrass approximation theorem \cite{kreyszig1978introductory}.

According to Taylor's expansion, if a real-valued function $f(t)$ is infinitely differentiable at a real number $t_0$, it could be decomposed into the sum of power series with the form of
\begin{equation}\label{eq:taylor-expansion}
  f(t) = f(t_0) + \frac{f^{(1)}(t_0)}{1!} (t - t_0) + ... + \frac{f^{(k)}(t_0)}{k!} (t - t_0)^k + ...,
\end{equation}
where $f^{(k)}(t_0)$ denotes the $k^{th}$-order derivative of $f(t)$ at $t_0$.

Note that (\ref{eq:taylor-expansion}) is also a polynomial with a special mathematical form rather than its general form below
\begin{equation}\label{eq:general-polynomial}
  f(t) = f_0 + f_1 t +  f_2 t^2 + ... + f_k t^k + ...,
\end{equation}
where $f_k$ are constant coefficients.

However, a function could be expanded as Taylor's series if and only if it is infinitely smooth, meaning infinitely differentiable. Thus we cannot directly apply the Taylor's series expansion for a general function $f(t)$ which may be discontinuous in (high-order) derivatives. To overcome this, we introduce a intermediate (temporary) function $p(t)$ as the Weierstrass approximation of $f(t)$. It means $p(t)$ is a polynomial with proper orders. Thus, we have $\forall \varepsilon >0 $, $\exists \bar K  > 0$, such that
\begin{equation}\label{eq:Weierstrass-approximation}
  \displaystyle \sup_{t}|f(t)-p_{\bar K}(t)| < \varepsilon,
\end{equation}
over a compact (i.e., closed in real space) interval, where $p_{\bar K}(t)$ denotes a polynomial with order of $\bar K$. For simplicity, we ignore $\bar K$ in notation. We have
\begin{equation}\label{eq:general-polynomial-with_p}
  p(t) = p_0 + p_1 t +  p_2 t^2 + ... + p_k t^k + ...,
\end{equation}

Thus, we could alternatively choose the polynomial in Taylor's form to regress $p(t)$ instead of $f(t)$ because only $p(t)$ is guaranteed to be infinitely differentiable. This will not lead to disaster, according to (\ref{eq:Weierstrass-approximation}), namely $p(t)$ and $f(t)$ are close enough.

Suppose we have interests in the properties at the discrete time index $n$, Eq. (\ref{eq:taylor-expansion}) could then be rewritten as (\ref{eq:taylor-expansion-discrete-time}).
\begin{equation}\label{eq:taylor-expansion-discrete-time}
  p(t) = p(n) + \frac{p^{(1)}(n)}{1!} (t - n) + ... + \frac{p^{(k)}(n)}{k!} (t - n)^k + ....
\end{equation}

\begin{figure}[htbp]
    \centering
    \subfigure[Global polynomial]{
        \begin{minipage}[htbp]{0.46\linewidth}
            \centering
            \includegraphics[height=3.85cm]{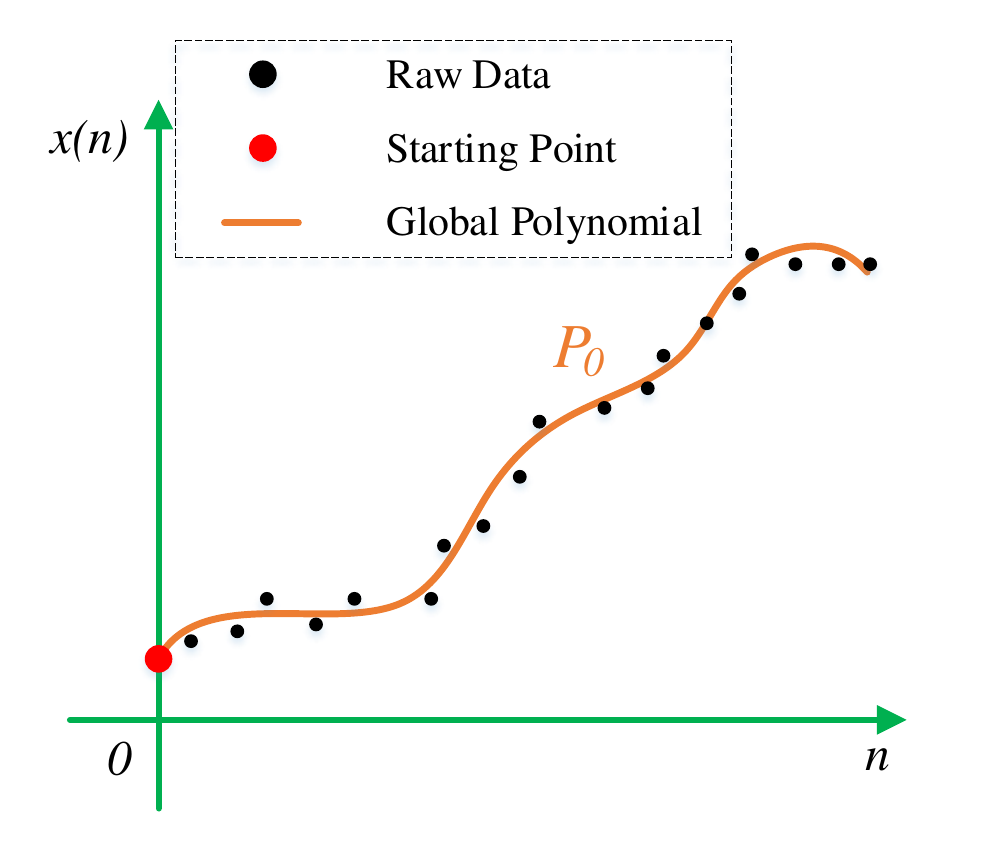}
        \end{minipage}
    }
    \subfigure[Local polynomial]{
        \begin{minipage}[htbp]{0.46\linewidth}
            \centering
            \includegraphics[height=3.85cm]{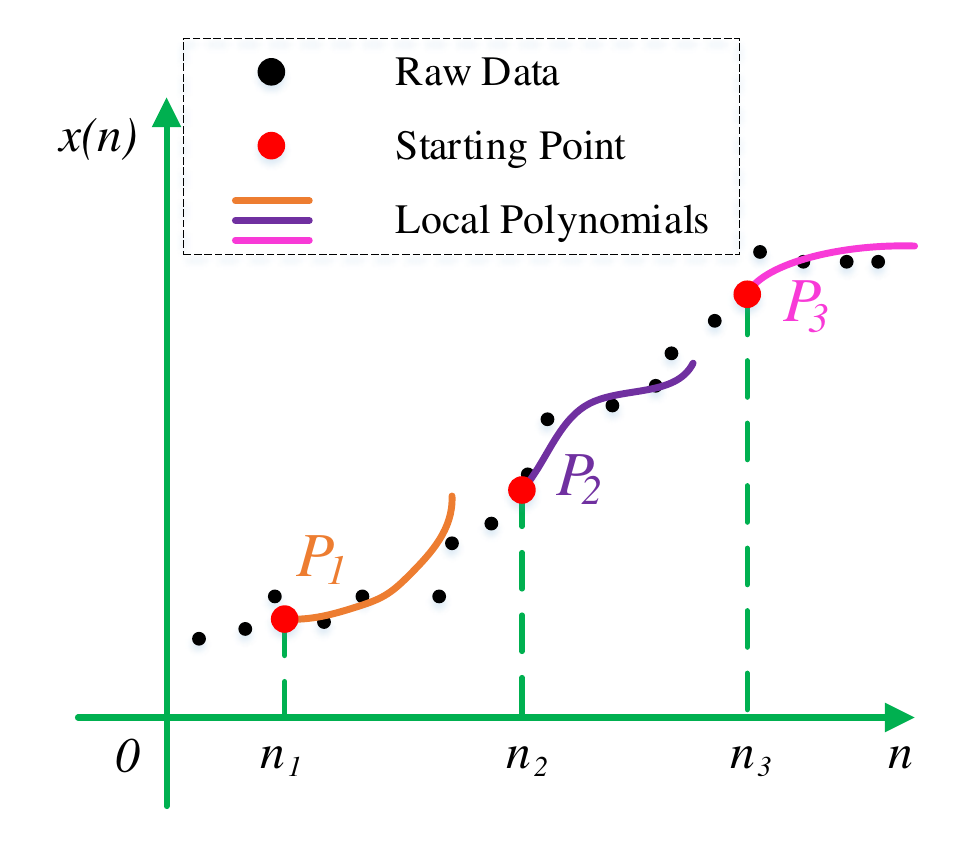}
        \end{minipage}
    }
    \caption{Global and local polynomial}
    \label{fig:local-global-polynomial}
\end{figure}

Thus, the traditional polynomial regression (\ref{eq:general-polynomial}) could be regarded as the special case of (\ref{eq:taylor-expansion}) when we investigate the problem from the starting point of the time, namely, $t_0 = 0$, meaning the polynomial (\ref{eq:taylor-expansion-discrete-time}) is a local polynomial, while (\ref{eq:general-polynomial}) is a global polynomial. For intuitive understanding, see Fig. \ref{fig:local-global-polynomial}.

If we only pay attention to the case of $t = n+1$, we have (\ref{eq:taylor-expansion-discrete-time}) as
\begin{equation}\label{eq:state-equation}
        p(n+1) 
               = \displaystyle \sum_{k = 0}^{\infty} \frac{p^{(k)}(n)}{k!} T^k = \displaystyle \sum_{k = 0}^{\infty} \frac{T^k}{k!} p^{(k)}(n).
\end{equation}

If we allow the truncation at the order of $K$, we have (\ref{eq:taylor-expansion-discrete-time}) and (\ref{eq:state-equation}) as
\begin{equation}\label{eq:state-equation-truncation}
        p(n+1) 
               = \displaystyle \sum_{k = 0}^{K} \frac{p^{(k)}(n)}{k!} T^k = \displaystyle \sum_{k = 0}^{K} \frac{T^k}{k!} p^{(k)}(n).
\end{equation}

\section{Understanding the Motion Modelling}
Let's study the natural philosophy, specially the motion description (modeling) of an macroscopic object in physics (in particular, in Theoretical Mechanics).

Suppose:
\begin{itemize}
  \item $s(n) := f(n)$ and $f^{(k)}(n) := 0,~ \forall k \geq 1$, we have (\ref{eq:state-equation}) as
        \begin{equation}\label{eq:static}
          s(n+1) = s(n);
        \end{equation}
  \item $v(n) := f^{(1)}(n)$ and $f^{(k)}(n) := 0,~ \forall k \geq 2$, we have (\ref{eq:state-equation}) as
        \begin{equation}\label{eq:velocity}
          s(n+1) = s(n) + v(n) T;
        \end{equation}
  \item $a(n) := f^{(2)}(n)$ and $f^{(k)}(n) := 0,~ \forall k \geq 3$, we have (\ref{eq:state-equation}) as
        \begin{equation}\label{eq:acceleration}
          s(n+1) = s(n) + v(n) T + \displaystyle \frac{1}{2} a(n) T^2.
        \end{equation}
\end{itemize}

When we use $s$ to denote the displacement, $v$ the velocity, and $a$ the acceleration of a moving object, we have the well-known kinematics equations (\ref{eq:static}) $\sim$ (\ref{eq:acceleration}) in physics. Note that $f^{(k)}(n) := 0,~ \forall k \geq 3$ admits that $a(n) := f^{(2)}(n)$ keeps constant over time. Thus we have Philosophy \ref{phi:motion} to disclose the nature of motion modeling. Before we have Philosophy \ref{phi:motion}, let's first have an axiom.

\begin{axiom}\label{axm:motion}
The motion of an physical macroscopic object is continuous, meaning it is impossible to see a position change of an object at a time moment. Thus none time elapse denotes none position change. This is concluded by (instant) energy and power limit of a moving macroscopic object.
\end{axiom}

Although the Axiom \ref{axm:motion} seems right intuitively and at least we cannot disprove it using any existing theoretical frames or experimental observations, we cannot assert it is absolutely right. Just as an axiom, let's suppose it indeed holds. Because this is the premise of the discussion of our Philosophy \ref{phi:motion}.

\begin{philosophy}\label{phi:motion}
Before the emergence of humans, the phenomena of motion already exist. When we try to understand the natural law of the motion, most of us get lost or started believing the existence of God (maybe the theism is correct, we do not hold standpoint here), until the great Newton was born. He found his Second Law and disclosed the relationship among the force, the mass and the acceleration. Further, he creatively used his concepts of integration and mathematically expressed the displacement with the velocity and the acceleration, which is thought to be the start of the scientific analysis and natural philosophy. Thus now, we should consider why Newton was right. What is underlying possibility of the correctness of this scientific frame? The viewpoints (or just thinking) of this note are as follows. According to Axiom \ref{axm:motion}, the motion law of an object could be represented by (\ref{eq:state-equation}) with sufficient order $K$ in any accuracy. The magic and interesting coincidence here is that the term $f^{(2)}(n)$ is proportional to the externally exerted force. Thus we assign a Philosophical concept to $f^{(2)}(n)$ and term as Acceleration. Subsequently, the other philosophical concepts like velocity and displacement were come up with and the relationships were bridged as (\ref{eq:static}) $\sim$ (\ref{eq:acceleration}), if the higher-order terms of (\ref{eq:state-equation}) are omitted.
\end{philosophy}

As a complement to Philosophy \ref{phi:motion}, we show Philosophy \ref{phi:is-motion-model-unique} 
below.

\begin{philosophy}\label{phi:is-motion-model-unique}
Some other issues should be raised. Is the time-variant local autocorrelated polynomial (TVLAP) model the unique one to explain the motion law of a moving object? Does it exist other approximation methods that are also plausible or interesting? Note that we do not know does not mean it does not exist. If exist, maybe we could have other philosophical concepts which are the conceptual counterparts of Acceleration and/or Velocity. For example, when we use Fourier series/transfrom to approximate the motion (changing pattern) of a signal \cite{chaparro2018signals}, or of a time series \cite{WANG2019ARIMA}, we have the concepts of Frequency rather than acceleration. The dominant difference here is whether and which parameters of the model we use could come across some physically existed concepts. For TVLAP, it is the pair of $f^{(2)}(n)$ and acceleration. For Fourier's, it is the pair of $\frac{k}{K T_s}$ and frequency. For more on $\frac{k}{K T_s}$, see Remark \ref{rem:Fourier}, Eq. (\ref{eq:FS-trig}) and \cite{WANG2019ARIMA}. Note also that the mathematics exists before we can realize them.
\end{philosophy}

\begin{remark}\label{rem:Fourier}
Let the operator $SIN(K|\bm A, \bm B)$ denote the sum of trigonometric functions, namely
    \begin{equation}\label{eq:FS-trig}
        \begin{array}{rll}
            f(n) &= A_0 + 2 \displaystyle \sum_{k=1}^{K} \left[ A_k \cos(\frac{2\pi}{K} kn) + B_k \sin(\frac{2\pi}{K} kn) \right] \\
                 &= \displaystyle \sum_{k=0}^{K} F_k e^{j\frac{2\pi}{K} kn} \\
                 &= \displaystyle \sum_{k=0}^{K} F_k e^{2\pi j\frac{k}{KT_s} t} \\
        \end{array}
    \end{equation}
    where $A_0$, $A_k$, $B_k$ and $F_k$ could be found in \cite{chaparro2018signals}, and $j$ is the complex unit. Note that $F_k$ here actually means the Discrete Fourier Series/Transform (DFS/DFT) coefficients of $f(n)$. $SIN(K|\bm A, \bm B)$ is shorted as $SIN(K)$. Note also that, by DFS/DFT theory, $SIN(K)$ could be any periodic time series with the period of $K$ \cite{WANG2019ARIMA}.
\end{remark}

\section{Results and Conclusion}
Interestingly and excitingly, the validity of above derivation could be supported by \cite{eager2016beyond,thompson2011snap}. In the mentioned literature of Physics, beyond the concepts of Velocity and Acceleration, the concepts of Jerk, Snap, Crackle and Pop are assigned to the high derivatives of $3$-order ($f^{(3)}(n)$), $4$-order ($f^{(4)}(n)$), $5$-order ($f^{(5)}(n)$) and $6$-order ($f^{(6)}(n)$), respectively.

The much higher order derivatives (orders larger than $6$) have not been studied in Physics, and we expect the advances in this direction.

As a closing note, we suggest a general kinematics equation
\begin{equation}\label{eq:general-kinematics}
        s(n+1) 
               = \displaystyle \sum_{k = 0}^{\infty} \frac{T^k}{k!} s^{(k)}(n),
\end{equation}
where $s$ denote the displacement of a moving object; $n$ the discrete time index.

\begin{remark}
  In Eq. (\ref{eq:general-kinematics}), if the displacement is not infinitely differentiable, then $s$ stand for the Weierstrass approximation of the displacement (instead of displacement itself).
\end{remark}

\begin{remark}
 If the external force ($F$) is infinitely differentiable, the high-order derivatives of displacement could be philosophically treated as the derivatives of external force. Note that:
 \begin{enumerate}
   \item $F = ma = m s^{(2)}$;
   \item $F^{(1)} = ma^{(1)} = m s^{(3)}$;
   \item $\cdots$;
   \item $F^{(k-2)} = ma^{(k-2)} = m s^{(k)},~k \ge 2$,
 \end{enumerate}
 where $m$ denote the mass of the object.
\end{remark}

\section*{Acknowledgment}
The corresponding author (S. Wang) would like to thank Dr. Xiaobo Li (Email: iselix@nus.edu.sg, National University of Singapore) for his insightful discussion.

\bibliographystyle{IEEEtran}
\bibliography{References}

\end{document}